\documentclass{ws-mpla}
\usepackage[super]{cite}
\usepackage{graphicx}
\PassOptionsToPackage{colorlinks=true,linkcolor=blue,citecolor=blue}{hyperref}
\def\lambdabar{{\mathchar'26\mkern-9mu\lambda}}

\begin{document}

\markboth{Carsten Henkel}
{Thermally Excited Quasiparticles}

\catchline{}{}{}{}{}

\title{\vspace*{-2ex}
THERMALLY EXCITED QUASIPARTICLES IN METALS,
DISPERSION FORCES, AND THE THERMAL ANOMALY
\vspace*{-2ex}}%

\author{\vspace*{-2ex}
\footnotesize CARSTEN HENKEL}

\address{Institute of Physics and Astronomy, University of Potsdam,
\\
Karl-Liebknecht-Str. 24/25,
14476 Potsdam, Germany
\\
henkel@uni-potsdam.de}

\maketitle



\vspace*{-2ex}
\begin{abstract}
We provide a brief review
of the contribution of thermally excited carriers to dispersion forces.
In a metal, these carriers generate
charge and current fluctuations whose spectral frequencies are
comparable to $k_B T/\hbar$. They
are very likely responsible for the ``plasma vs. Drude'' anomaly.
\keywords{Dispersion force; quasi-particles.}
\ccode{PACS Nos.: 42.50.Pq, 42.50.Lc, 44.40.+a}
\end{abstract}

\section{Dispersion Forces at Different Distances and Relevant Frequencies}

There is a long history behind the temperature dependence of dispersion
forces.
The physical problem is easy to
formulate: since the pioneering work of Casimir\cite{Casimir_1948} who
switched from the viewpoint of van der Waals forces, generated by
dipole fluctuations of material particles,
to the viewpoint of field modes and their zero-point energy, it is
natural to ask how thermally excited states of
matter or of field modes participate in the momentum exchange
between two objects.

For a given physical system it is relatively easy to decide whether
thermal occupation plays a role or not: within
the \emph{frequency spectrum} of the relevant modes, those with
$\hbar\omega \gg k_B T$ are essentially in the quantum regime and
contribute with their zero-point energies, while frequencies
$\hbar\omega \ll k_B T$ are enhanced
by a factor scaling with $k_B T / \hbar\omega$ at
low frequencies.
Since most experiments are performed
at room temperature $T \sim 300\,\mathrm{K}$,
one gets the characteristic far-infrared value
$\omega_T/2\pi = k_B T / (2\pi\hbar) \sim 26\,\mathrm{meV}
= 6.2\,\mathrm{THz}
= 208\,\mathrm{cm}^{-1}$.

The relevant frequency spectrum is not
trivial to find because dispersion forces
are naturally generated by a continuum of frequencies
-- remember the theory of Eisenschitz and London for
the van der Waals force
with its integration over virtual transitions reaching
even into the ionisation continuum\cite{Eisenschitz_1930}. In
the Casimir effect between parallel plates, an electrodynamic
scale emerges naturally, namely the
cutoff frequency for a cavity of length $d$,
$\omega_d = \pi c / d$. The
distinction between ``quantum'' and ``thermal'' regimes
thus maps into short
and large distances (see Table~\ref{t:regimes} and Fig.\,\ref{fig:Mehra67}),
separated quite accurately by the scale
$\lambdabar_T = c / 2 \omega_T$,
giving at room temperature
$\lambdabar_T \sim 3.8\,\mu\mathrm{m}$. This sets a challenge
for experiments with macroscopic bodies because dispersion forces
are weak, and relatively accurate control over the geometry (parallelism
of plates) is required. The scaling of the Casimir free energy
$\mathcal{F}$
per unit area between two large plates can be found
by dimensional analysis in three regimes, as shown in Table~\ref{t:regimes}.

\begin{table}[b!]
\tbl{Distance regimes and Casimir free energy\newline between two
metallic plates. See text for definitions.}
{\begin{tabular}{@{}ccc@{}}
\toprule
\multicolumn{2}{c}{quantum = short distance: $d \ll \lambdabar_T$}
&
thermal = large distance: $d \gg \lambdabar_T$
\\
\colrule
non-retarded: $d \ll \lambdabar_p$
&
retarded: $\lambdabar_p \ll d$
&
\\
\colrule
$\displaystyle
\mathcal{F}
\sim - \frac{ \hbar \omega_p }{ d^2 }$
& $\displaystyle
\mathcal{F}
\sim - \frac{ \hbar c }{ d^3 }$
& $\displaystyle
\mathcal{F}
\sim - \frac{ k_B T }{ d^2 }$
\\[1.5ex]
\botrule
\end{tabular}\label{t:regimes}}
\end{table}
Experiments at closer distances benefit from larger forces,
and one expects them to be dominated by the zero-point fluctuations
of matter and field. It thus came as a surprise that
in 2000, Bostr\"om and Sernelius
reported significant thermal
corrections to the Casimir pressure between metallic plates at distances
smaller than the thermal wavelength.\cite{Bostrom_2000b}
At about the same time, Svetovoy and Lokhanin worked on precision
calculations and also had to take into
account both temperature and material
parameters.\cite{Svetovoy_2000b,Svetovoy_2001} This period saw the
birth of the ``thermal anomaly'' in the Casimir effect.
By the anomaly, we mean the unusually large thermal correction
that already appears in the intermediate range
$\lambdabar_p \lesssim d \ll \lambdabar_T$.
After the
experimental work by the groups of Lamoreaux and of Mohideen,
additional experiments were
reported\cite{Lamoreaux_1997a,Mohideen_1998,Harris_2000,Bressi_2002,Decca_2003a,Chen_2004},
and a comparison to theory at the few-percent level seemed within
reach. In earlier periods, theorists checked their calculations
by considering the special case of a cavity with perfectly conducting
boundaries. This case had been studied for example
by Mehra\cite{Mehra_1967}
and by Schwinger, DeRaad, and Milton\cite{Schwinger_1978} 
as a limiting case.
Schwinger and co-workers provided a critical survey of asymptotic expressions
worked out by Lifshitz\cite{Lifshitz_1956}, in particular for the 
finite-temperature case:
\begin{quote}
[...] we have verified Lifshitz' formula\cite{Lifshitz_1956} for the Casimir
force between parallel dielectric interfaces, including the temperature
dependence. Where the Russian calculations went wrong\cite{Hargreaves_1965}
was in their specialization to metal plates. A careful reading of their papers
shows that, through relatively trivial errors, they obtained incorrect results
for the Casimir force, for perfect conductors at finite temperatures, and for
imperfect conductors at zero temperature.
\end{quote}
In this paper, we review the physical origin of the
unexpected temperature dependence of the dispersion force between
imperfect conductors.

\section{Lifshitz Theory and Thermal Casimir Pressure Correction}

The theoretical work after Lifshitz confirmed the validity of his
celebrated formula for the Casimir pressure $P$ between two plates.
We copy it here for plates~1 and~2 as an integral over real frequencies
\begin{equation}
P(d, T) = \int_0^\infty\!\frac{ {\rm d}\omega }{ 2\pi }\,
\big[ \tfrac{1}{2} + \bar{n}( \omega, T) \big]
\int_{{\sf L}}\!\frac{ k_z {\rm d}k_z }{ 2\pi }
\mathrm{Re}\bigg(
2 \hbar k_z \sum_{\sigma \,=\, \mathrm{p,s}}
\frac{ r_{1\sigma} r_{2\sigma} \,{\rm e}^{ 2 {\rm i} k_z d } }{ 1 - r_{1\sigma} r_{2\sigma} \,{\rm e}^{ 2 {\rm i} k_z d } }
\bigg),
\label{eq:Lifshitz55}
\end{equation}
where $\bar{n}( \omega, T)$ is the Bose distribution,
providing a natural cutoff $\omega \lesssim \omega_T$ for the
thermal contribution; the $k_z$-integral
starts at $k_z = \omega/c$ and goes via the origin to
$k_z = +{\rm i} \infty$ (symbolically written ${\sf L}$).
In particular the
works by Kats and Bimonte made it clear that the
$r_{\sigma}$ are indeed reflection amplitudes for
electromagnetic waves in the p (TM) and s (TE)
polarisation.\cite{Kats_1977,Bimonte_2007b} 
The same expression applies for zero-point
and thermal fluctuations, apart from the different frequency ranges that
dominate the two contributions.
Expanding in powers of $r_{1\sigma} r_{2\sigma}$, we get a multiple
reflection sum, the basis for the
scattering approach to the Casimir
pressure.\cite{Reynaud_2010}
Perfect reflectors give $r_{\rm s} = -1$ and
$r_{\rm p} = +1$,
and the denominator
constrains the integral to cavity modes with $k_z = n \pi / d$;
imaginary $k_z$ (evanescent waves) then do not contribute.

\begin{figure}[h]
\vspace*{-03mm}
\centerline{%
\includegraphics[height=4.5cm]{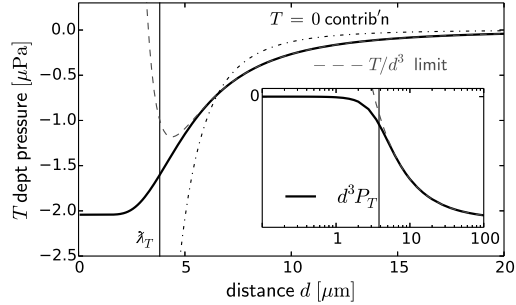}
}
\vspace*{-0.2cm}
\caption[]{%
Temperature-dependent Ca\-simir pressure between two per\-fectly
conducting plates.\cite{Mehra_1967}. The gray dashed line
is the difference between the high-temperature $T/d^3$ and the zero-temperature
$\hbar c / d^4$ asymptotes. Inset: pressure multiplied by $d^3$.
Negative values correspond to an attractive force.
\protect\label{fig:Mehra67}}
\end{figure}

For illustration, we plot
in Fig.\,\ref{fig:Mehra67} the difference $P(d, T) - P(d, 0)$
of the Casimir pressure between perfectly conducting plates.%
\cite{Mehra_1967}
The thick line is the exact result, the dot-dashed line gives
for comparison the zero-temperature pressure
$P(d, 0)$, 
the gray dashed line the
high-temperature asymptote
$P(d, T) \simeq - \zeta(3) k_B T / ( 4\pi d^3 )$ that dominates
for $d \gg \lambdabar_T$, as expected.
At smaller
distances, it is, for perfect reflectors, a tiny correction, since
at $d = 1\,\mu\mathrm{m}$, $P(d, 0) = -1.3\,\mathrm{mPa}$.

Any realistic material brings in a characteristic cutoff beyond which
it is transparent, as Casimir noted himself.
The reflection coefficients encode these material properties,
as they must,
since electromagnetic vacuum fluctuations mani\-fest themselves
only in so far they couple to matter.\cite{Jaffe_2005,Nikolic_2016}
For gold, the cutoff is set by the plasma frequency
$\omega_p / 2\pi \simeq 9\,\mathrm{eV}$,
as computed from the free carrier density.
The plasma wavelength
$\lambdabar_p = c / \omega_p \sim 22\,\mathrm{nm}$
(also known as Meissner penetration depth)
is not a microscopic length compared to the closest
distances $d \sim 100\,\mathrm{nm}$ achieved
in recent Casimir force experiments.
The regime $d \ll \lambdabar_p$ is quantum rather than thermal, non-retarded
and amenable to the standard approximations of theoretical 
chemistry, even though van der Waals interactions need special
tools like the adiabatic connection theorem
to be reached with density functional 
techniques, for example.\cite{DiStasio_2014, Woods_2016}. 

\section{Nature and Contribution of Matter-Dominated Modes}

What is the impact of the material parameters on the Casimir force?
This question has been at the heart of the debate around its
temperature dependence. It is related to the question what kind
of quasi-particles may contribute to the force, in addition to
the electromagnetic field.
Looking at reflection coefficients, a non-ideal medium broadens
the resonances that would be infinitely sharp between two perfect conductors,
and makes them disappear above the plasma (or transparency) edge.
Also new types of electromagnetic quasi-particles appear,
like surface plasmon polaritons, located at $\omega_p/\sqrt{2}$
(and below) at a metal-vacuum interface.\cite{Barton_1979,Raether_book} Their
hybridization provides a simple way to understand the Casimir
pressure in terms of ``bonding'' and ``anti-bonding
modes'', leading to the scaling
$\mathcal{F} \sim
\hbar \omega_p / d^2$ (Table~\ref{t:regimes}).%
\cite{Henkel_2004a,Volokitin_2004,Intravaia_2005}
Surface polaritons being bound to the metal surface,
they correspond to purely imaginary wave vectors $k_z$;
this `leg' of the $\mathrm{d}k_z$-integral
in Eq.\,(\ref{eq:Lifshitz55}) (evanescent waves in the TM polarization)
thus becomes dominant at distances $d \sim \lambdabar_p$.

{%
It has been argued that electric field fluctuations of very low frequencies inside a metal cannot be relevant for the Casimir force because of the fast response of the charge density on its boundaries.%
\cite{Geyer_2007}
}%
The estimate is based on Ohm's law
${\bf j} = \sigma {\bf E}$ that gives, combining with charge conservation
and Coulomb's law
\begin{equation}
\partial_t \rho = - \nabla \cdot \sigma {\bf E}
= - \frac{ \sigma }{ \varepsilon_0 } \rho
\,.
\label{eq:wrong-charge-relaxation}
\end{equation}
{%
This would give a time scale $\varepsilon_0 / \sigma \sim
10^{-19}\,\mathrm{s}$ (for gold) after which the total electric field
in plates of finite size has decayed to zero.
One may raise the question whether for such a fast decay,
the DC conductivity is still meaningful.
}%
Indeed, a different time scale is found from Drude's model for the current
density
\begin{equation}
\partial_t {\bf j} + \frac{ {\bf j} }{ \tau } =
\frac{ \sigma }{ \tau } {\bf E}
\label{eq:Drude-model}
\end{equation}
with a relaxation time $\tau$.
Taking the divergence leads to the telegraphist's equation
\begin{equation}
\partial^2_t \rho + \frac{ \partial_t \rho }{ \tau } +
\frac{ \sigma }{ \varepsilon_0 \tau } \rho = 0
\,,
\label{eq:telegraphist}
\end{equation}
that yields an eigenfrequency with imaginary part
$- 1/(2\tau)$: the charge density
cannot relax faster than the current density. These processes cannot
be irrelevant because their rate is quite comparable
to the room-temperature characteristic frequency $\omega_T$: a typical
value for gold is $\tau \sim 27\,\mathrm{fs}$ so that
$1/(2\pi\tau)
\sim 24\,\mathrm{meV}
= 5.9\,\mathrm{THz}
= 190\,\mathrm{cm}^{-1}$.
{%
(With respect to spatial scales, one may raise doubts about
the relevance of the length scale
$\varepsilon_0 c / \sigma \sim 1\,\mathrm{\AA}$, that has led to the claim
that the conductivity should be irrelevant for macroscopic distances
involved in typical Casimir forces.\cite{Svetovoy_2001})
}%

The Drude conductivity that follows from Eq.\,(\ref{eq:Drude-model})
\begin{equation}
\sigma( \omega ) = \frac{ \sigma }{ 1 - {\rm i} \omega \tau }
\,,
\label{eq:Drude-sigma}
\end{equation}
has been discussed in many papers and conferences.%
\cite{Mostepanenko_2006a, QFExt07, Klimchitskaya_2009, Klimchitskaya_2009a,
QFExt09, Milton_2012c, Brevik_2014, Milton_2017a, Bimonte_2017a, Henkel_2017a}
It has been argued
that for Casimir calculations, the lossless variant, the ``plasma model''
\begin{equation}
\sigma( \omega ) =
\frac{ {\rm i} \sigma }{ \omega \tau }
=
\frac{ {\rm i} \varepsilon_0 \omega_p^2 }{ \omega }
\,,
\label{eq:plasma-sigma}
\end{equation}
should be used,
where the Drude parameters combine into the plasma frequency, excluding
actual Ohmic losses. This is not so obvious for the thermal correction
that arises from $\omega \ll \omega_T \sim 1/\tau$ where
Eq.\,(\ref{eq:plasma-sigma}) is a poor approximation.

We are touching here those modes to which the \emph{thermal anomaly} can be
attributed:
it was clear from the beginning that they are TE-polarized and appear
at low frequencies.%
\cite{Bostrom_2000b,Svetovoy_2001,Torgerson_2004,Bimonte_2006c,Torgerson_2006a}.
An inspection of Eq.\,(\ref{eq:Lifshitz55}) shows that they appear for
imaginary $k_z$ (evanescent waves).
In Ref.\,\refcite{Svetovoy_2006a}, it was explicitly stated that these
modes correspond to magnetic fields.
This insight also provides a physical explanation why they are not
efficiently reflected by a metallic surface: by the
Bohr--von Leeuwen theorem, a metal is transparent to static magnetic
fields, unless quantum effects like superconductivity play a
role.\cite{Bimonte_2009b}
The consequences of this behaviour can be seen directly in the large-distance
behaviour of the Casimir free energy computed within classical
field theory:\cite{Buenzli_2005,Jancovici_2005a} the power law
$\mathcal{F} \sim k_B T / d^2$ comes with a prefactor that is half as large
compared to the result obtained with perfectly reflecting bodies.
Also the plasma model~(\ref{eq:plasma-sigma})
yields a non-zero static result for the reflection coefficient $r_{\rm s}$
when the conventional Fresnel formula is used:
\begin{equation}
r_{\rm s} =
\frac{ k_{z} - k_{zm} }{ k_{z} + k_{zm} }
\,,\qquad
k_{zm}
= \sqrt{ {\rm i} \mu_0 \omega \sigma( \omega ) + k_z^2 }
= \sqrt{ k_z^2 - 1/\lambdabar_p^2 }
\,.
\label{eq:plasma-rs}
\end{equation}
This has led Intravaia and the present author to the statement that
the imaginary conductivity of Eq.\,(\ref{eq:plasma-sigma}) actually
describes a superconductor with its distinct feature of the Meissner
effect. (It is obvious, of course, that this statement is only true
at low frequencies. The conductivity of a BCS superconductor becomes
complex at frequencies above the gap, for example.)

\section{Contribution of Diffusing Magnetic Fields}

\begin{figure}[t!]
\vspace*{-0.3cm}
\centerline{%
\includegraphics[height=4.5cm]{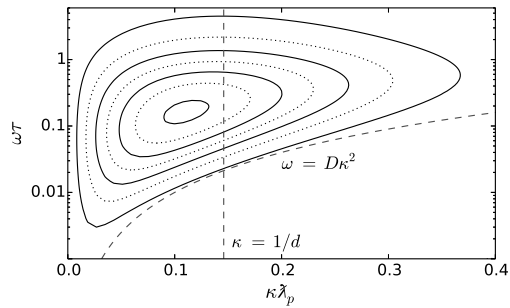}
}
\vspace*{-0.2cm}
\caption[]{Mode-resolved Casimir pressure between two Drude metals
as a function of imaginary wave vector $k_z = {\rm i}\,\kappa$ and
real frequency $\omega$ (see main text for details).
Typical Drude parameters for gold ($\omega_p\tau \approx 390$,
magnetic diffusivity $D \approx 0.018\,\mathrm{m^2/s}$)
and distance $d = 150\,\mathrm{nm} = 6.9\,\lambdabar_p$.
\protect\label{fig:TE-DOS}}
\end{figure}

If a normal conductor is penetrable to low-frequency magnetic fields,
how is it possible that these fields contribute to the Casimir effect?
They are evanescent and contribute to the Lifshitz formula along
imaginary $k_z = {\rm i} \kappa$. Using the Drude conductivity
in the Fresnel formula then yields instead of Eq.\,(\ref{eq:plasma-rs})
($\mathop{\mathrm{Im}} k_{zm} \ge 0$)
\begin{equation}
r_{\rm s} =
\frac{ {\rm i} \kappa - k_{zm} }{ {\rm i} \kappa + k_{zm} }
\,,\qquad
k_{zm}
= \sqrt{ \frac{ {\rm i} \omega / D }{ 1 - {\rm i} \omega \tau }
- \kappa^2 }
\,,
\qquad
D = \frac{ 1 }{ \mu_0 \sigma } = \frac{ \lambdabar_p^2 }{ \tau }
\,.
\label{eq:Drude-rs}
\end{equation}
Here, the quantity $D$
is the \emph{magnetic diffusivity} that
describes how time-dependent magnetic fields penetrate into a
normal conductor (a standard non-invasive technique for material
inspection called eddy current testing).
This class of modes is illustrated in Fig.\,\ref{fig:TE-DOS},
where
the real part of the
last parenthesis in Eq.\,(\ref{eq:Lifshitz55}) is shown,
evaluated for $k_z = {\rm i}\,\kappa$ and multiplied by $\kappa$.
The vertical line gives
the natural cutoff $\kappa = 1/d$.
One notes that the peak (central contour) is located
near $\omega \sim 1/\tau$.

The thermal anomaly in the Casimir force thus arises from thermal
current fluctuations that diffuse across
a metallic body, partially cross its surface and couple in the form of
evanescent magnetic fields to another metallic body.
These modes are therefore not surface, but bulk ``modes'', and
have a peculiar status because their eigenfrequencies
are purely imaginary, $\omega_k = -{\rm i} D k^2$.
For $k \lambdabar_p \ll 1$, this scaling law provides
a natural upper limit for magnetic bulk fields, as shown in dashed
in Fig.\,\ref{fig:TE-DOS}, projected onto real frequencies.
In thermal equilibrium,
the damping of fields applied from outside the body is exactly
compensated by the fields generated from thermal fluctuating currents
by Ampere's law.\cite{Ford_1993a,Rosa_2010}
Ingold and co-workers have worked
out in detail the free energy of an overdamped particle
with an imaginary eigenfrequency.\cite{Milton_2017a,Ingold_2009b,Haake_1985}
It also shows
thermal anomalies like a negative specific heat or entropy.
These arise from the subtraction of the heat bath 
coupled to the overdamped particle. 

The quantum field theory of overdamped magnetic field modes has also
been worked out.%
\cite{Intravaia_2009a, Henkel_2009a, Bordag_2011a, Bordag_2011b}
It has been shown that the relevant electromagnetic
mode density in the two-plate cavity (easy to read off from Eq.\,(\ref{eq:Lifshitz55})) is essentially exhausted in spectral weight
for $\omega \lesssim 1/\tau$ by diffusive magnetic fields.
This has clarified the negative
entropies of the Casimir problem:%
\cite{Svetovoy_2003a,Bezerra_2004}
they correspond to a tiny, surface-dependent entropy shift of
magnetic fields diffusing throughout the bulk.\cite{Intravaia_2009a, Henkel_2009a}

In discussions of the thermal anomaly, media with
temperature-dependent scattering times
have been put forward. For a perfect crystal, one
expects $\tau(T) \to \infty$ as $T \to 0$ because all scattering
mechanisms freeze out. The Casimir entropy has been worked out in
this limit\cite{Intravaia_2009a, Henkel_2009a}
and found to be of the order of $S \sim - k_B / d^2$.
This was claimed a violation of the Nernst heat theorem that tells us
that the entropy of an isolated body should vanish at absolute
zero.\cite{Bezerra_2004}
For this specific case, magnetic diffusion could resolve
the paradox:\cite{Henkel_2009a} when taking the limit, the
frequency range $\omega \lesssim 1/\tau(T) < \omega_T$
where these modes can be
excited moves to ever lower frequency, but always stays in the
thermal regime. This is an example of a subtle interplay of taking
the zero-temperature or zero-frequency limits.\cite{Ellingsen_2008b}
The perfect crystal thus degenerates into a ``glassy state'' filled
with current loops interlocked with magnetic field lines. This theoretical
example of a perfect (rather than super\mbox{-)}conductor contains a
surface-dependent disorder entropy because, as mentioned above,
magnetic fields in the two bodies can couple to each other via the
vacuum gap.

\section{Conclusion}

To conclude, the challenge remains to explain the discrepancy
between Lifshitz theory for non-perfect conductors
and micromechanical experiments in Decca's group%
\cite{Decca_2003b,Decca_2005,Decca_2007a, Bimonte_2016}
and atomic force microscope experiments in Mohideen's group%
\cite{Chang_2012,Banishev_2013,Liu_2019}.
Theoretical limits
like the perfect conductor or the Nernst theorem are of little help,
since one has to deal with a real system here. We conjecture that surface
roughness may yield surprises for 
the particular class of 
low-frequency
magnetic near fields (in the $\mathrm{THz}$ band and on
the $10{-}100\,\mathrm{nm}$ spatial scale) that is 
not yet directly amenable
to spectroscopic experiments.

\section*{Acknowledgments}

This work has been supported by the \emph{Deutsche Forschungsgemeinschaft}
through the DIP program (grant nos. Schm 1049/7-1 and Fo 703/2-1).

\end{document}